\documentclass[prl,nobibnotes,altaffilletter,amsmath,amssymb,amsfonts]{revtex4}
\usepackage[latin1]{inputenc}
\usepackage{graphicx}
\usepackage{amsmath}
\usepackage{latexsym}
\usepackage{epsfig}

\begin{document}
\title{`Isomorphs' in liquid state diagrams}
\author{Nicoletta Gnan, Thomas B. Schr{\o}der, Ulf R. Pedersen, Nicholas P. Bailey, and Jeppe C. Dyre}
\affiliation{DNRF centre  ``Glass and Time,'' IMFUFA, Department of Sciences, Roskilde University, Postbox 260, DK-4000 Roskilde, Denmark}
\date{\today}
\begin{abstract}
A liquid is termed strongly correlating if its virial and potential energy thermal equilibrium fluctuations in the NVT ensemble are more than 90\% correlated [Phys. Rev. Lett. {\bf 100}, 015701 (2008)]. The fluctuations of a strongly correlating liquid are well approximated by those of an inverse power-law intermolecular potential. Building on this fact we here define ``isomorphic lines'' in the state diagram of a strongly correlating liquid. It is shown from computer simulations of the Kob-Andersen binary Lennard-Jones liquid that no aging is associated with jumps between two isomorphic points. Isomorphic state points have the same excess entropy, the same reduced average relaxation time, the same dynamics, and the same scaled radial distribution functions. Finally we calculate the equation for isomorphs in the virial / potential energy diagram for Lennard-Jones type liquids and show that all such 12-6 liquids have the same isomorphs; these may be scaled into a master isomorph.
\end{abstract}

\maketitle

\section{Introduction}

How much does knowledge of a system's thermal equilibrium fluctuations at one state point tell us about its behavior at other state points? In principle, complete knowledge of the fluctuations provides enough information to determine the partition function and thus the density of states, from which the free energy at other state points may be calculated. In practice, only second order moments of the fluctuations may be determined reliably. These generally give little knowledge of the system's properties away from the state point in question. It was recently shown that van der Waals type liquids are ``strongly correlating'' in the sense defined below \cite{ped08a,ped08b,bai08a,bai08b,cos08}; this implies that such liquids have a ``hidden'' approximate scale invariance \cite{sch08b}. Because of this, important global information about the system may be obtained from knowledge of the virial and potential energy second-order moments at one state point. This unusual situation in statistical mechanics is the focus of the present note.

In order to briefly recapitulate the definition of the class of strongly correlating liquids, recall that the pressure $p$ is a sum of the ideal gas term $Nk_BT/V$ and a term reflecting the interactions, $W/V$, where $W$ is the so-called virial:

\begin{equation}\label{Wdef}
pV \,=\,
Nk_BT+W\,.
\end{equation}
This equation is usually thought of as describing thermodynamic averages, but it also describes the instantaneous fluctuations. The instantaneous ideal-gas pressure term is a function of the particle momenta. The instantaneous virial $W$ is a function of the particle positions, $W=W({\bf r}_1,..., {\bf r}_N)$. In the same way, of course, the energy is a sum of the kinetic energy and the potential energy $U$. If $\Delta U$ is the (instantaneous) potential energy minus its thermodynamic average and $\Delta W$ the same for the virial, at any given state point the $WU$ correlation coefficient $R$ is defined by (where sharp brackets denote equilibrium NVT ensemble averages)

\begin{equation}\label{R}
R \,=\,
\frac{\langle\Delta W\Delta U\rangle}
{\sqrt{\langle(\Delta W)^2\rangle\langle(\Delta U)^2\rangle}}\,.
\end{equation}
We define ``strongly correlating liquids'' by requiring $R>0.9$ \cite{ped08a}.

Strongly correlating liquids include \cite{ped08a,ped08b,cos08,bai08a} the standard Lennard-Jones (LJ) liquid (and the classical LJ crystal), the Kob-Andersen binary LJ liquid as well as other binary LJ type mixtures, a dumbbell-type liquid of two different LJ spheres with fixed bond length, a system with exponential repulsion, a seven-site united-atom toluene model, the Lewis-Wahnstr{\"o}m OTP model, and an attractive square-well binary model. Liquids that are not strongly correlating include water and methanol \cite{bai08a}. The physical understanding developed in Refs. \cite{bai08a,bai08b} is that strong $WU$ correlations are a  property of van der Waals liquids and some or all metallic liquids. Liquids with directional bonding like covalent and hydrogen-bonding liquids do not have strong $WU$ correlations. Likewise, ionic liquids are not expected to be strongly correlating because of the different distance dependence of their short-range repulsions and the Coulomb interactions -- competing interactions spoil the correlations.

Strongly correlating liquids appear to have simpler physics than liquids in general, an observation that has particular significance for the highly viscous phase \cite{gt_rev}. It has been shown that supercritical (experimental) argon is strongly correlating \cite{ped08a,bai08b}, that strongly correlating viscous liquids have all eight frequency-dependent thermoviscoelastic response functions \cite{ell07,chr08} given in terms of just one \cite{ped08b} (are ``single-parameter liquids,'' i.e., have dynamic Prigogine-Defay ratio close to unity \cite{bai08b}), that strongly correlating viscous liquids obey density scaling, i.e., that their relaxation time $\tau$ depends on density $\rho$ and temperature as $\tau= F(\rho^\gamma/T)$ \cite{density_scaling_exp}, and that even complex systems like a biomembrane may exhibit significant $WU$ correlations for their slow degrees of freedom \cite{ped08c}. The property of strong virial / potential energy correlation is maintained even out of equilibrium and during crystallization \cite{sch09}.

Whenever equilibrium fluctuations of the virial are plotted versus those of the potential energy for a strongly correlating liquid, an elongated ellipse appears \cite{ped08a,cos08,bai08a}. The ``slope'' $\gamma$ of this ellipse in the $WU$ diagram is given by (NVT averages)

\begin{equation}\label{gamma}
\gamma \,=\,
\sqrt{
\frac{\langle(\Delta W)^2 \rangle }{\langle(\Delta U)^2 \rangle}}\,.
\end{equation}
This quantity, which is weakly state-point dependent, is the number entering into the density scaling relation $\tau= F(\rho^\gamma/T)$ \cite{sch08b,cos09}. Thus for strongly correlating liquids knowledge of equilibrium fluctuations at one state point provides a prediction about how the relaxation time varies with density and temperature.

What causes strong $WU$ correlations? A hint comes from the well-known fact that an inverse power-law pair potential, $v(r)\propto r^{-n}$ where $r$ is the distance between two particles \cite{ipl}, implies perfect correlation \cite{ped08a,bai08b}. In this case $\gamma=n/3$. In simulations of the standard LJ liquid we found $\gamma\cong 6$ which corresponds to $n\cong 18$ \cite{ped08a}. Although this may seem puzzling at first sight given the expression defining the LJ potential, $v_{LJ}(r)=4\epsilon[(r/\sigma)^{-12}-(r/\sigma)^{-6}]$, if one fits the repulsive part of the LJ potential by an inverse power law, an exponent $n\cong 18$ is required \cite{ped08a,bai08b,ben03}. This is because the attractive $r^{-6}$ term makes the repulsion steeper than the bare repulsive $r^{-12}$ term would imply. 

Reference \cite{bai08b} gave a thorough discussion of the correlations with a focus on the standard single-component LJ liquid, including also a treatment of the classical crystal where $0.99<R<1$ at low temperature. According to Ref. \cite{bai08b} the $r$-dependent effective exponent $n$ which controls the correlation is not simply that coming from fitting the repulsive part of the potential, but rather $n^{(2)}(r)\equiv -2-rv'''(r)/v''(r)$. This number is approximately 18 around the LJ minimum. In fact, the LJ potential may here be fitted very well with an ``extended'' inverse power-law potential \cite{bai08b}, $v_{\rm LJ}(r)\cong A r^{-n}+B+Cr$ with $n\cong 18$; for this particular potential of course $n^{(2)}(r)=n$. At constant volume the linear term contributes little to the virial and potential-energy fluctuations: When one nearest-neighbor interatomic distance increases, another decreases in such a way that the sum is almost constant. Thus almost correct Boltzmann probability factors are arrived at by using the inverse power-law approximation, implying that thermal fluctations at one state point are well described by this approximation. The equation of state, however, is poorly represented by the inverse power-law (IPL) approximation \cite{bai08b}, and for instance the IPL relation $W=\gamma U$ does not apply for liquids that are not 100\% correlating.

Inspired by the results summarized above, in this note we demonstrate a number of results based on one single assumption, the existence of curves in the phase diagram of a strongly correlating liquid on which the state points have configurations that are statistically equivalent in the canonical ensemble. Such curves will be referred to as \emph{isomorphs}.

\section{Isomorph definition and predicted properties}

Two state points (1) and (2) with temperatures $T_1$ and $T_2$ and densities $\rho_1$ and $\rho_2$, respectively, are said to be \emph{isomorphic} if they obey the following: Any two configurations of state points (1) and (2), $({\bf r}_1^{(1)}, ... , {\bf r}_N^{(1)})$ and $({\bf r}_1^{(2)}, ... , {\bf r}_N^{(2)})$, which may be scaled into one another, i.e. obey

\begin{equation}\label{eq:relation_1}
\rho_1^{1/3}[\mathbf{r}_i^{(1)}-\mathbf{r}_j^{(1)}]\,=\,\rho_2^{1/3}[\mathbf{r}_i^{(2)}-\mathbf{r}_j^{(2)}]\,\,(i,j=1,...N)\,,
\end{equation}
have proportional Boltzmann statistical weights:

\begin{equation}\label{eq:rel2}
e^{-U({\bf r}_1^{(1)}, ... , {\bf r}_N^{(1)})/k_BT_1}\, =\, C_{12}e^{-U({\bf r}_1^{(2)}, ... , {\bf r}_N^{(2)})/k_BT_2}\,.
\end{equation}
Here $U({\bf r}_1, ... , {\bf r}_N)$ is the potential energy function and $C_{12}$ is a constant, which in general depends on both state points. An IPL liquid obeys Eq. (\ref{eq:rel2}) with $C_{12}=1$ for states that obey $\rho_1^{n/3}/T_1=\rho_2^{n/3}/T_2$. No systems except IPL liquids obey Eq. (\ref{eq:rel2}) rigorously. Nevertheless, based on computer simulations we show below that the existence of isomorphs is a good approximation, thus providing new insight into the physics of strongly correlating liquids. Before presenting the simulation results we briefly summarize some predicted properties of isomorphs, leaving their more detailed justification to a future publication:

\begin{enumerate}

\item {\it Two isomorphic states have the same excess entropy:} $S_{{\rm ex},1}=S_{{\rm ex},2}$. The excess entropy is by definition the entropy in excess to that of an ideal gas at same volume and temperature. This quantity is always negative. The excess entropy equals $-k_B\int P({\bf R})\ln P({\bf R})d{\bf R}/V^N$ where $P({\bf R})$ is the normalized canonical configuration space probability. Equation (\ref{eq:rel2}) implies that isomorphic points have the same normalized canonical configuration space probability, thus the same excess entropy. In other words, isomorphs are configurational adiabats. In particular, the specific heat along an isomorph equals the ideal gas specific heat.

\item {\it Isomorphic states have the same average reduced relaxation time and the same reduced diffusion constant (i.e., they lie on the same``isochrone'')}. This prediction applies whether the dynamics is Brownian or given by Newton's equations. It follows from the fact that the potential energy landscapes of two isomorphic state points are identical, except for an overall scaling that is compensated by the adjusted temperature and an additive constant.

\item {\it Isomorphic states have the same (reduced) dynamics, e.g., the same relaxation functions.} This follows from the fact that their scaled potential-energy landscapes are identical. Property 3 implies property 2, of course.

\item {\it No aging is associated with jumps between two isomorphic state points.} If density and temperature are changed instantaneously from one state point to values characterizing a new state point which is isomorphic with the initial state, the system will instantaneously be in equilibrium. This is because all normalized Boltzmann probability factors are identical for the two systems. Thus isomorphs are predicted to be ``wormholes'' between which one can jump instantaneously from equilibrium to equilibrium, even when the states themselves are characterized by long relaxation times.

\item {\it Isomorphic states have the same scaled radial distribution functions.} This follows from the fact that their normalized Boltzmann probability factors are identical.

\end{enumerate}

We emphasize once again that no systems except IPL liquids have state points that are exactly isomorphic. Thus whenever state points are referred to as as isomorphic, more correctly one should say that they are {\it approximately} isomorphic. Note that properties 2 and 3 are consistent with the intriguing discovery by Ngai, Casalini, Capaccioli, Paluch, and Roland \cite{nga05} that for varying pressure and temperature states with the same relaxation time have the same shape of the alpha relaxation dielectric loss peak (this should only to apply for strongly correlating liquids, for instance not for hydrogen-bonding liquids).

The concept of isomorphs may shed light on the excess-entropy scaling proposed long ago by Rosenfeld, according to which the dimensionless transport coefficients (diffusion constant, viscosity, thermal conductivity, etc) of dense fluids are all functions of the configurational entropy \cite{rosenfeld}. For any strongly correlating liquids this property follows from Eq. (\ref{eq:rel2}), but we cannot explain Rosenfeld scaling of dense fluids generally (see also Ref. \cite{truskett}).

Recall that density scaling is the fact that the reduced relaxation time is a function of $\rho^{\gamma}/T$ \cite{density_scaling_exp}. The exponent $\gamma$ entering into this expression is given by Eq. (\ref{gamma}) \cite{sch08b,cos09}. This makes it easy to identify approximate isomorphs in computer simulations of strongly correlating liquids: In the simulations presented below two state points are regarded as isomorphic in the simulations below if $\rho_1^{\gamma}/T_1=\rho_2^{\gamma}/T_2$.

\section{Results from computer simulations}

The well-known Kob-Andersen 80:20 binary Lennard-Jones liquid (KA) \cite{KA} is a nice model system that is easily supercooled. This system is a strongly correlating liquid \cite{ped08a,bai08a}. The slope $\gamma$ varies slightly with state point, but at low and moderate pressures and temperatures this quantity is always close to 6. 

\begin{figure}
\begin{center}
\includegraphics[width=10cm]{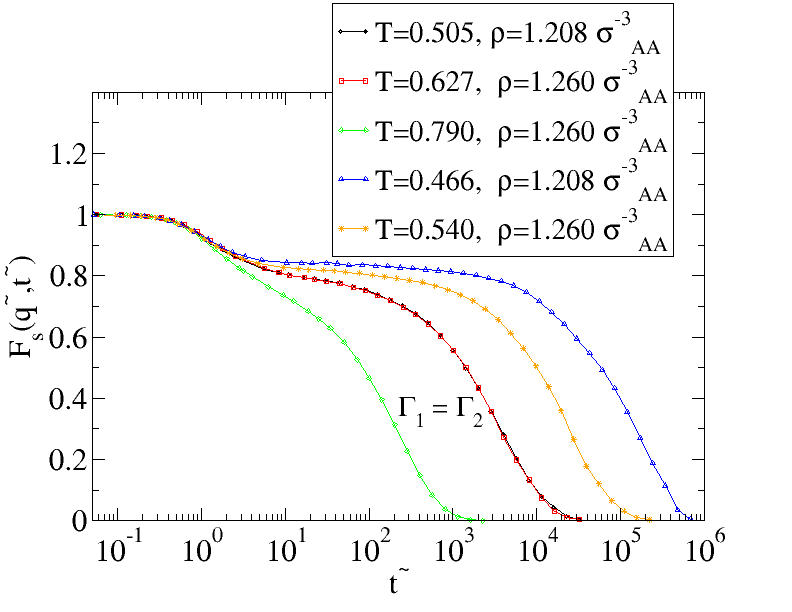}
\end{center}
\caption{The self part of the intermediate scattering functions $F_s(\tilde{q},t)$ at the wavevector corresponding to the first peak of the structure factor as function of the reduced time $\tilde t$ for five state points of the KA liquid with 8000 particles. The two isomorphic points (i.e., with the same $\Gamma=\rho^\gamma/T$) have identical scaled self part of the intermediate scattering functions.}\label{figure1}
\end{figure} 

Figure \ref{figure1}  reports results from molecular dynamics simulations of the KA mixture with $N=8000$ particles. The figure shows the self part of the intermediate scattering functions $F_s(\tilde q,\tilde t)$ at the wavevector corresponding to the first peak of the structure factor for five state points as functions of reduced time (i.e., time scaled by $\rho^{-1/3}(k_BT/m)^{-1/2}$). The simulations were performed in the NVT ensemble using the Nose-Hoover thermostat with characteristic time equal to 0.5 in MD units. Two of the five state points in the figure are isomorphic. These two points not only have the same average relaxation time, but the same relaxation behavior including the short-time ``cage-rattling'' contributions. This figure confirms isomorph properties 2 and 3.

\begin{figure} 
\begin{center}
\includegraphics[width=12cm]{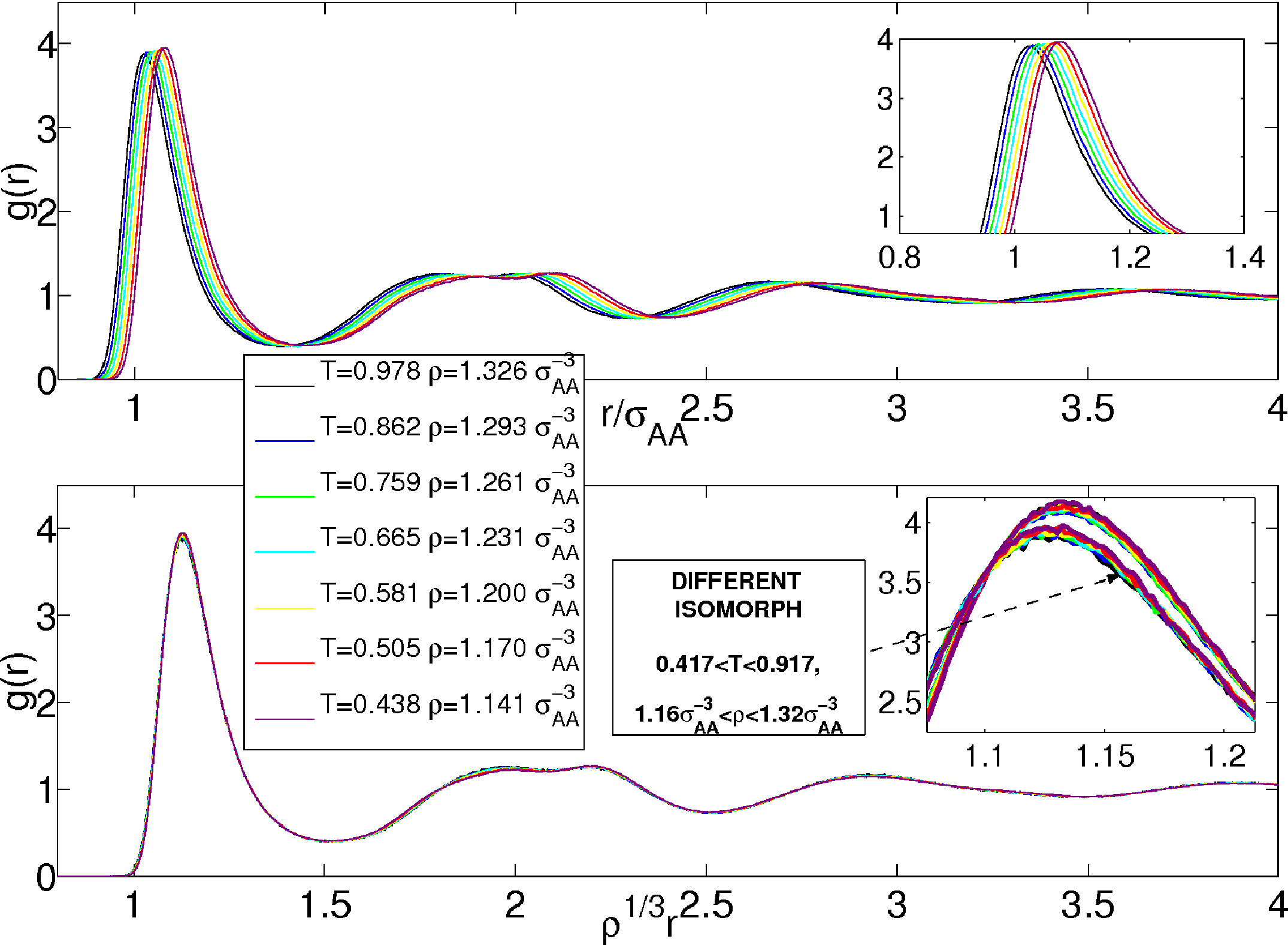}
\includegraphics[width=10cm]{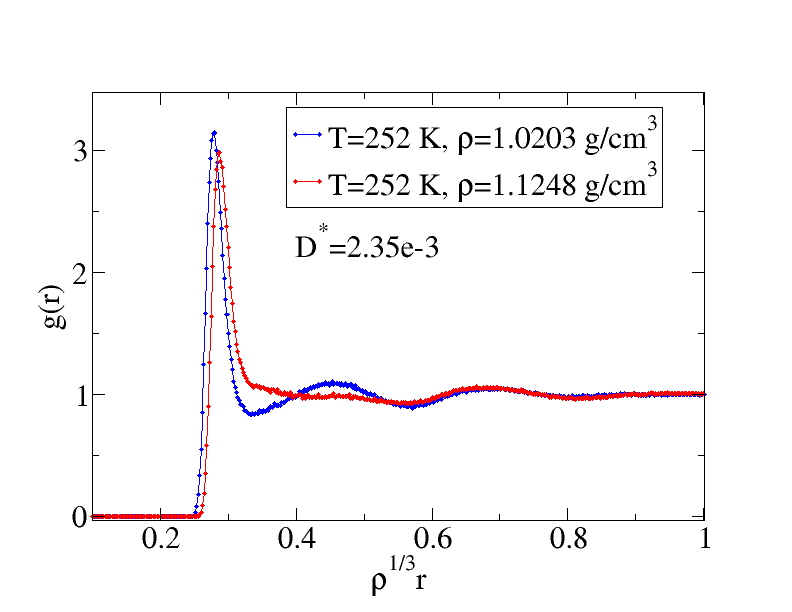}
\end{center}
\caption{(a) AA particle radial distribution function for the KA system at different isomorphic state points. The inset focuses on the first peak. 
(b) The same radial distribution functions, now as functions of the reduced distance, showing a collapse. The inset, which focuses on the first peak, shows further results for another collection of isomorphic state points.
(c) Radial distribution function as functions of the reduced distance for two state points of SPC water, which is not a strongly correlating liquid. The two state points have the same scaled relaxation time; thus if water had isomorphs the two radial distribution functions should collapse.}\label{fig:figure2}
\end{figure} 

Figure \ref{fig:figure2}(a) gives AA radial distribution functions for various isomorphic state points of the KA system, where A is the large particle. Temperature varies by more than a factor of two. Nevertheless, Fig. \ref{fig:figure2}(b) shows that when plotted as function of the reduced distance ($\rho^{1/3}r$), there is good data collapse. This confirms isomorph property 5. The inset of (b) focuses on the radial distribution function peak, adding data for a second collection of isomorphic state points. Finally, Fig. \ref{fig:figure2}(c) shows two radial distribution functions for water, our favorite non-strongly correlating liquid -- water's correlation coefficient $R$ is close to zero, a fact that reflects the existence of a density maximum. The two water state points have the same reduced diffusion constant; thus if water had isomorphs, the two radial distribution functions should be identical except for an overall scaling. This is not the case.

\begin{figure} 
\begin{center}
\includegraphics[width=9cm]{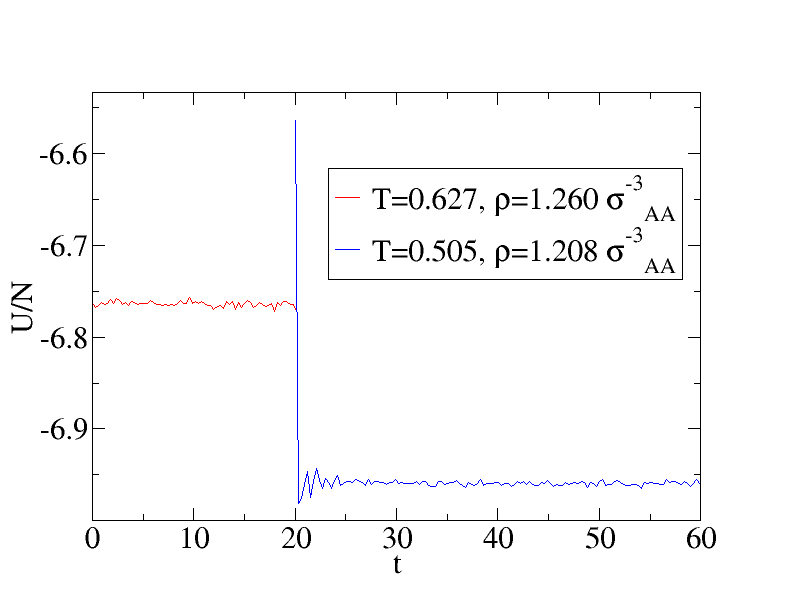}
\includegraphics[width=9cm]{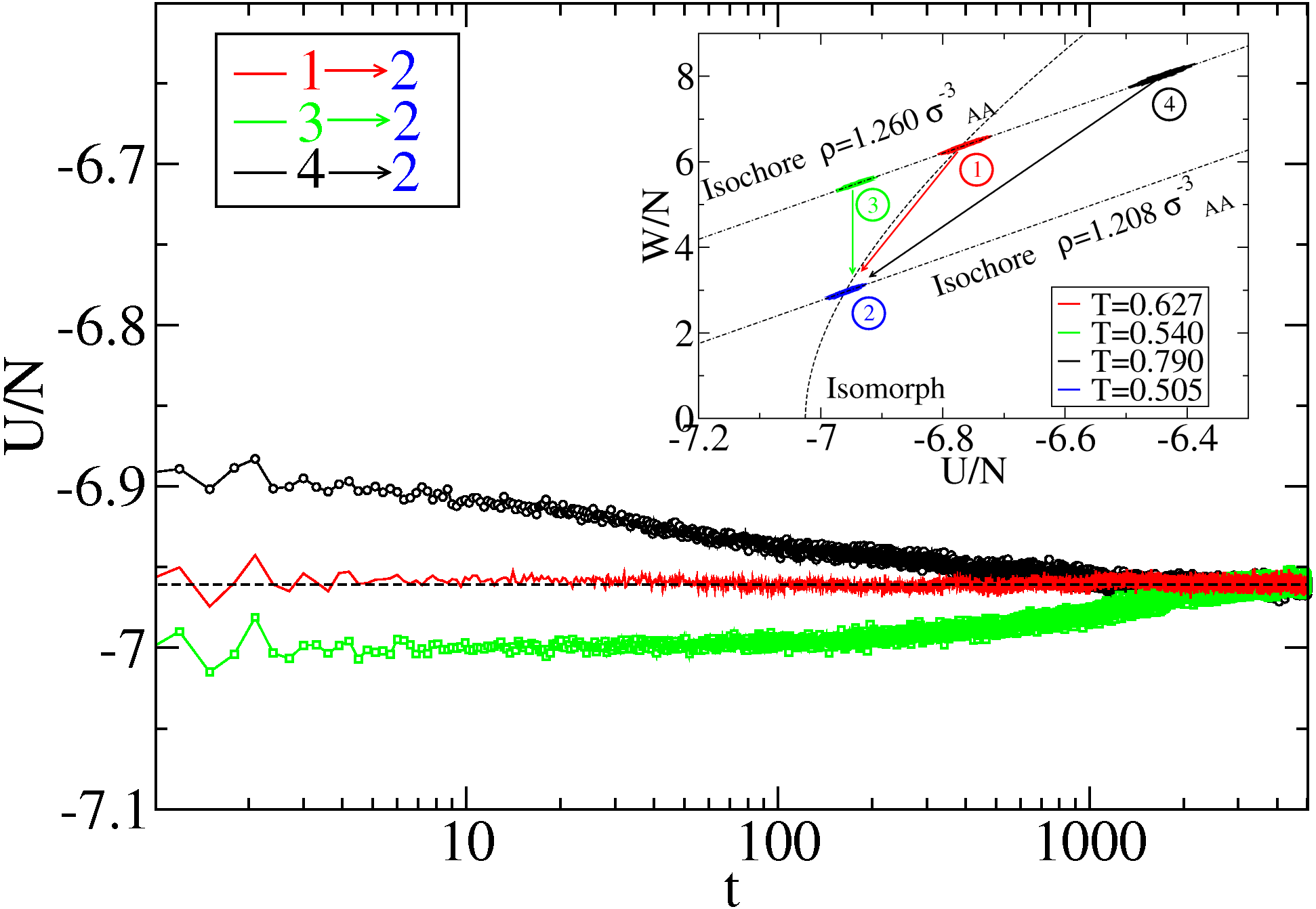}
\includegraphics[width=9cm]{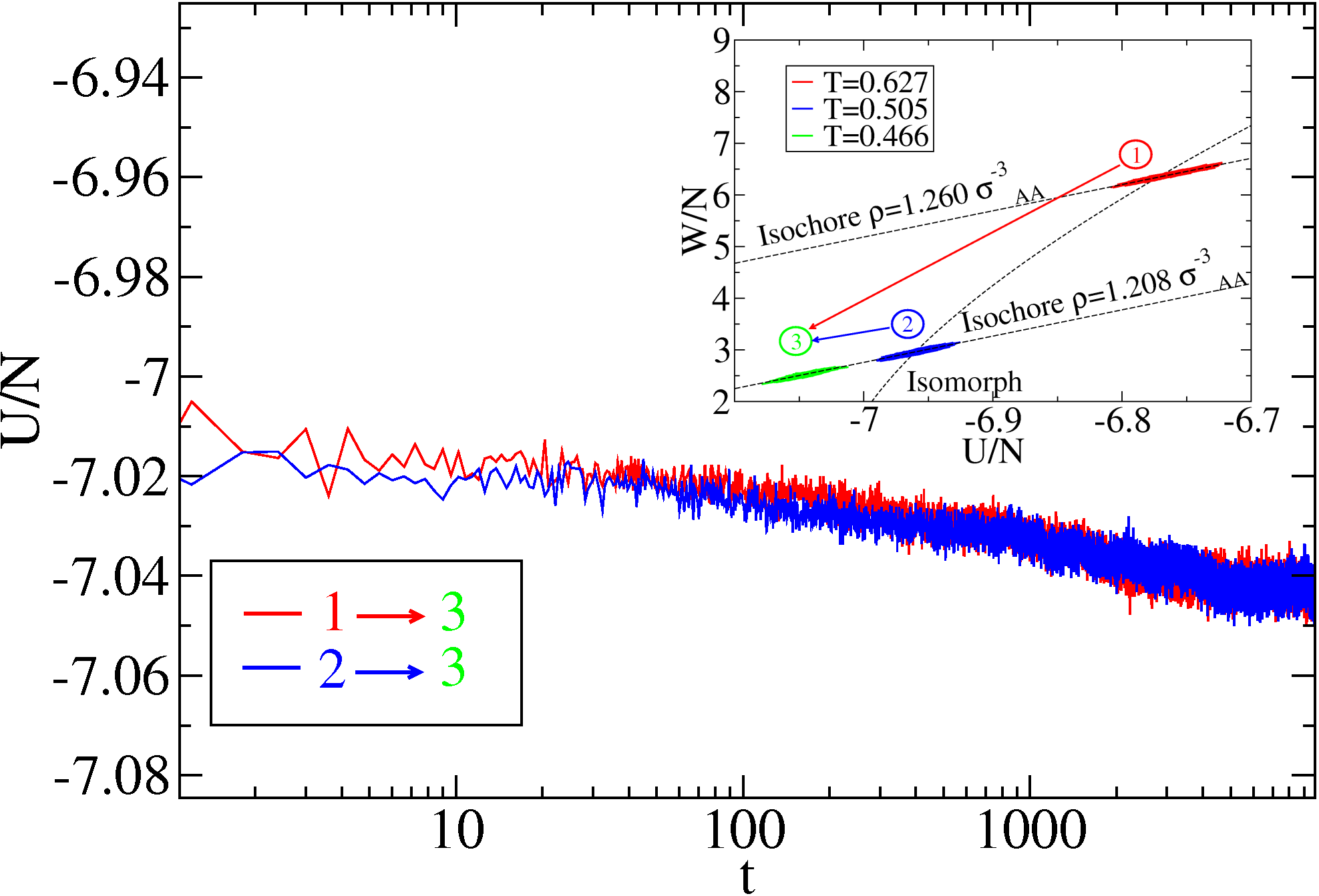}
\end{center}
\caption{(a) Results from simulating an instantaneous temperature and density jump applied to the KA system, bringing it to a state that is isomorphic with the initial state. Except for the very fast transient there is no relaxation associated with the jump, showing that the system is instantaneously in equilibrium. The relaxation time of the state points is roughly 500 in reduced units (they have the same relaxation time).
(b) Results of the potential energy relaxation towards equilibrium for the KA system comparing the jumps $1\rightarrow 2$, $3\rightarrow 2$, and $4\rightarrow 2$. The $1\rightarrow 2$ jump is between isomorphic state points and thus predicted not to involve any change beyond the immediate change (compare (a)); the two other jumps are not between isomorphic state points.
(c) Results of the potential energy relaxation towards equilibrium for the KA system comparing the jumps $1\rightarrow 3$ and $2\rightarrow 3$. The two relaxations are predicted to be identical.}\label{fig:figure3}
\end{figure}

Figure \ref{fig:figure3}(a) shows the time evolution of the potential energy when a temperature and density jump is made on a KA system in thermal equilibrium. The jump brings the system to a new state point that is isomorphic with the initial point (the state points in question are the two isomorphic state points of Fig. 1).  The jump was obtained in the following way: We instantaneously increased the box length without changing any particle positions (the overshoot is due to this) and simultaneously changed the thermostat temperature to the final temperature. There is no sign of relaxation towards equilibrium after the jump. Thus the system is instantaneously in equilibrium after the jump, as predicted for jumps between isomorphic state points (property 4). 

Figure \ref{fig:figure3}(b) extends on the point made in (a) that a jump between two isomorphic state points conserves thermal equilibrium ($1\rightarrow 2$). In contrast, the $3\rightarrow 2$ and $4\rightarrow 2$ jumps both have slow relaxations to equilibrium after the jump was applied at $t=0$. The inset is a scatter plot of virial vs potential energy for the three different starting points (same volume) and for state point (2), which is isomorphic with state point (1) but has a different volume. The curves show relaxation towards the equilibrium value represented by the dashed line, the curve that oscillates around the average value from the beginning shows the $1\rightarrow 2$ jump. Again, we see that instantaneous equilibration is a feature of jumps between isomorphic points.

The preservation of equilibrium for jumps between state points on an isomorph has consequences that are relevant for generic aging experiments. Consider the inset of Fig. \ref{fig:figure3} (c). Suppose we start in equilibrium at state (1) and change temperature and volume to $T_3$ and $V_3$. State (2) has volume $V_3$ and lies on the same isomorph as state (1). Now suppose that instead of jumping directly from point (1) to point (3), one first jumps from (1) to (2) and then immediately after to point (3). In that case, whenever relaxation times are long, the system never ``discovers'' that it spent a short amount of time in state (2) (where it immediately equilibrated). In other words, the relaxation behavior going from (2) to (3) must be the same as for the (1) to (3) experiment. Figure \ref{fig:figure3} (c) confirms this.

\section{The equation for Lennard-Jones isomorphs in the $WU$ state diagram}

\begin{figure} 
\begin{center}
\includegraphics[width=9cm]{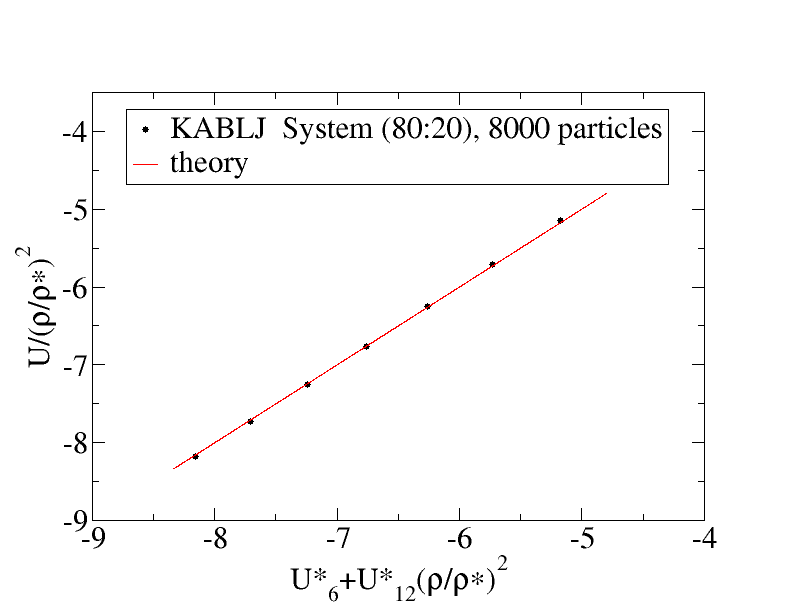}
\includegraphics[width=9cm]{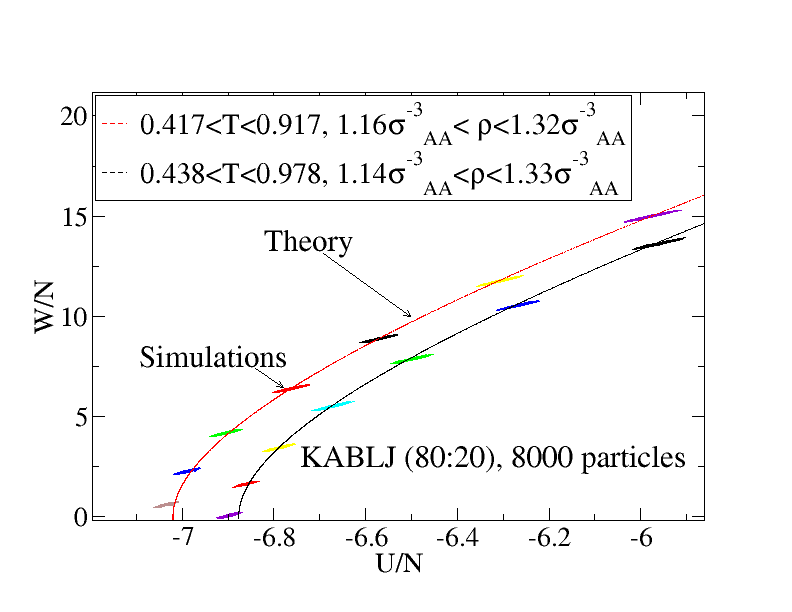}
\includegraphics[width=10cm]{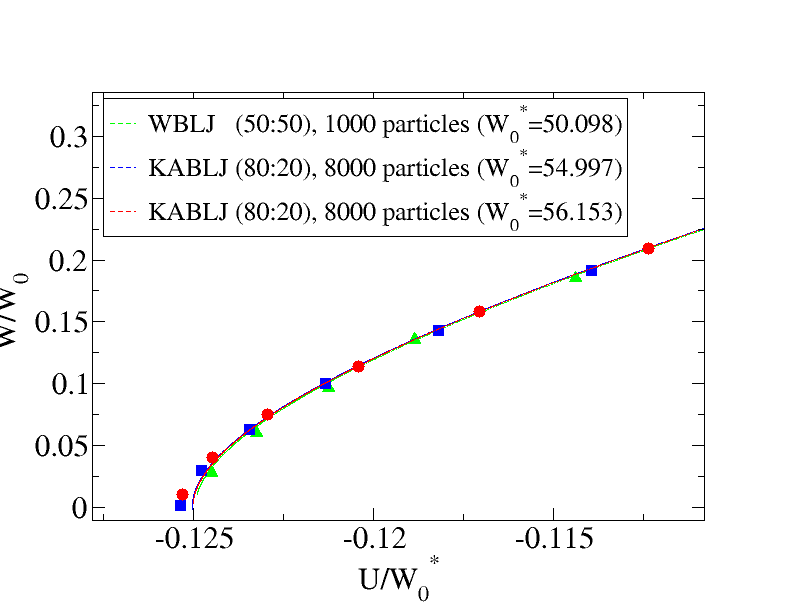}
\end{center}
\caption{
(a) Check of Eq. (\ref{isoU}).
(b) Two isomorph predictions according to  Eqs. (\ref{isoU}) and (\ref{isoW}) compared to two series of state points (marked by blobs) that each have the same reduced relaxation time.
(c) The ``master isomorph'', collapsing two KA isomorphs and one for the Wahnstr{\"o}m binary LJ liquid by scaling with $W_0^*$ defined as the virial on the same isomorph when $U=0$.}
\label{fig:figure4}
\end{figure} 

Is it possible to say anything general about the isomorphic lines in the $WU$ diagram? Consider a system characterized by an arbitrary number of $12$-$6$ LJ interactions:

\begin{equation}\label{generalLJ}
 \phi_{ij}(r_{ij})=\phi_{ij}^{(12)}(r_{ij})+\phi_{ij}^{(6)}(r_{ij})
\end{equation}
where $\phi_{ij}^{(m)}(r_{ij})=\varepsilon_{ij}^{(m)} [\sigma_{ij}^{(m)}/{r_{ij}}]^m $. For the total potential energy we get

\begin{equation}\label{generalU}
 U=U_{12}+U_{6},\qquad	U_m\equiv \sum_{i<j}\phi_{ij}^{(m)}(r_{ij})\,.
\end{equation}
The definition of the virial implies that

\begin{equation}\label{generalW}
 W=4U_{12}+2U_{6}\,.
\end{equation}
Denoting a reference point by ``*'', if we define the relative density ${\tilde{\rho}} \equiv \rho/\rho^*$, one has at points that are isomorphic wiht the reference point, because the scaled radial disbributions functions are identical,

\begin{equation}\label{isoU}
 U=U_{12}^{*} {\tilde{\rho}}^4+U_{6}^{*}{\tilde{\rho}}^2
\end{equation}
and
\begin{equation}\label{isoW}
 W=4U_{12}^{*} {\tilde{\rho}}^4+2U_{6}^{*} {\tilde{\rho}}^2\,.
\end{equation}
These two equations parametrize the isomorph curves in the $(U,W)$-plot for any $12$-$6$ LJ system. In particular, they imply that along an isomorph $dU=(W/ {\tilde{\rho}}) d{\tilde{\rho}}$, which leads to $dU=Wd\ln{\tilde\rho}=Wd\ln\rho=-Wd\ln V=-p_c dV$ (where $p_c=W/V$ is the configurational pressure). This expresses the fact that an isomorph is an excess entropy adiabat (property 1). The two parameters $U_{12}^{*}$ and $U_{6}^{*}$ can be calculated by solving Eqs. (\ref{generalU}) and (\ref{generalW}) at the reference point. Note that the predicted isomorphs are independent of the parameters characterizing the potential. Thus all kinds of different 12-6 LJ-type liquids are predicted to have the same isomorphs (provided an identical reference point exists).

Figure \ref{fig:figure4} compares the above to simulations. In Fig. \ref{fig:figure4} (a) Eq. (\ref{isoU}) is tested; the red line in this plot has slope $1$ indicating that $U/\tilde{\rho}^2 = U_6^*+U_{12}^*\tilde{\rho}^2$ as predicted. Figure \ref{fig:figure4} (b) shows two predicted isomorphs according to Eqs. (\ref{isoU}) and (\ref{isoW}) in the $WU$ diagram (dotted lines) compared to two collections of isomorphic state points. The agreement is good, with some deviations only when $W\cong 0$. Finally, Fig. \ref{fig:figure4} (c) shows that a ``master isomorph'' is obtained by scaling both $U$ and $W$ by $W_0^*$ defined as the virial for the point on the same isomorph with $U=0$. We here also give data from simulations of the Wahnstr{\"o}m binary LJ liquid, which has quite different properties from the KA system. All systems collapse to the master isomorph -- a parabola which is identified as follows. Equations (\ref{isoU}) and (\ref{isoW}) imply ${\tilde{\rho}}^2=(4U-W)/(4U^*-W^*)$ and ${\tilde{\rho}}^4=(W-2U)/(W^*-2U)$. The requirement ${\tilde{\rho}}^4=({\tilde{\rho}}^2)^2$ implies $[(4U-W)/(4U^*-W^*)]^2=(W-2U)/W^*-2U^*)$. Since $W_0^*=(4U-W)^2/(W-2U)$, in terms of the variables $X\equiv U/W_0^*$ and $Y\equiv W/W_0^*$ the master isomorph equation is $(4X-Y)^2=Y-2X$.

\section{Executive summary}

Strongly correlating liquids have lines in their state diagram -- ``isomorphs'' -- with the property that different points on an isomorph have potential energy landscapes that to a good approximation may be scaled into one another. A number of static and dynamic properties are invariant along an isomorph. An isomorph acts as wormhole in the state diagram.

\end{document}